\documentclass[twocolumn,superscriptaddress,showpacs,amsmath,amssymb,aps,prl]{revtex4-1}
\usepackage{graphicx}
\usepackage{amsmath}
\usepackage{multirow}

\newcommand{\Ntl}{n_{\mathrm{U}}^\leftarrow}
\newcommand{\Ntr}{n_{\mathrm{U}}^\rightarrow}
\newcommand{\Ntu}{n_{\mathrm{U}}^\uparrow}
\newcommand{\Ntd}{n_{\mathrm{U}}^\downarrow}
\newcommand{\Nbl}{n_{\mathrm{L}}^\leftarrow}
\newcommand{\Nbr}{n_{\mathrm{L}}^\rightarrow}
\newcommand{\Nbu}{n_{\mathrm{L}}^\uparrow}
\newcommand{\Nbd}{n_{\mathrm{L}}^\downarrow}

\begin{document}

\title{Path-integral Monte Carlo method for the local $\mathbf Z_2$ Berry phase}

\author{Yuichi Motoyama}
\affiliation{Department of Applied Physics, University of Tokyo, Tokyo 113-8656, Japan}
\author{Synge Todo}
\affiliation{Institute for Solid State Physics, University of Tokyo, 7-1-26-R501 Port Island South, Kobe 650-0047, Japan}

\date{\today}

\begin{abstract}
We present a loop cluster algorithm Monte Carlo method for
calculating the local $Z_2$ Berry phase of the quantum spin models.
The Berry connection, which is given as the inner product of two
ground states with different local twist angles, is expressed as a
Monte Carlo average on the worldlines with fixed spin configurations
at the imaginary-time boundaries.  The ``complex weight problem''
caused by the local twist is solved by adopting the meron cluster
algorithm.  We present the results of simulation on the antiferromagnetic
Heisenberg model on an out-of-phase bond-alternating ladder to demonstrate that our
method successfully detects the change in the valence bond pattern at the quantum phase transition point.  We
also propose that the gauge-fixed local Berry connection can be an
effective tool to estimate precisely the quantum critical point.
\end{abstract}


\maketitle

In low-dimensional systems or frustrated magnets, strong quantum fluctuations often disrupt the classical long-range orders.
For example, the ground state of the $S=1$ antiferromagnetic Heisenberg (AFH) chain is
not the classical N\'{e}el state but the Haldane state~\cite{Haldane1982}, which is non-magnetic ($S_{\text{tot}}=0$) and gapful.
This state is understood in terms of the valence bond solid (VBS) picture~\cite{AffleckKLT1987}. The ground state of the $S=1/2$ AFH ladder is another spin-gap state -- the spin dimer state.
While these models exhibit quantum phase transitions between a spin-gap state and another,
these states are purely quantum state and thus they cannot be characterized by any classical order parameters.
Such states and phase transitions have been studied for decades by using topological order parameters.
In spin systems, the string order parameter~\cite{NijsR1989} and its generalizations~\cite{Oshikawa1992,TodoMYT2001,NakamuraT2002a}
have been used to describe the Haldane and the dimer phases.
Although the string order parameters have been used widely especially in numerical studies, they work only in one-dimensional systems.

The local $Z_2$ Berry phase, recently proposed by Hatsugai~\cite{Hatsugai2006}, is one of the topological order parameters.
This is a phase that the ground state wave function acquires under the twist of a bond on the lattice.
The value of the local $Z_2$ Berry phase is defined on each bond, plaquette, and so on,
and quantized to $0$ or $\pi$ (mod $2\pi$) when the system has a finite gap and an antiunitary symmetry such as the time reversal symmetry.
This order parameter is stable against any perturbation unless the gap collapses.
Therefore, one can catch the quantum critical points by seeing the change of the spatial pattern of the Berry phases
even at finite-size simulations.
In the AFH models, for example,
the value of the local Berry phase on a bond is $\pi$ or $0$
depending on whether or not a singlet spin pair exists on the bond, respectively.
Other models that have been studied by this order parameter are itinerant electron systems~\cite{MaruyamaH2007} and Kondo singlets~\cite{MaruyamaH2009}.
It is considered that this topological order parameter can be applied to two- or higher-dimensional systems as well as one-dimensional ones.
The Berry phase and the local $Z_2$ Berry phase are related~\cite{RyuH2006, HiranoKH2008} to the topological entanglement entropy,
which is another tool to detect the nontrivial topological phase~\cite{KitaevP2006, LevinW2006}.

The local $Z_2$ Berry phase, however, has been calculated numerically only by the exact diagonalization 
and thus the simulation size was strongly restricted (in $S=1/2$ AFH models, up to about $40$ spins).
To verify the conjecture, that is, the local $Z_2$ Berry phase also works for higher-dimensional systems,
and to apply the order parameter for attacking non-trivial quantum phase transitions,
an unbiased numerical method, such as the path-integral quantum Monte Carlo method (PIQMC)~\cite{Suzuki1976},
which can treat larger systems, is essential.

The PIQMC calculation for the Berry phase has so far not been done since this has two obstacles.
The first obstacle is that the Berry connection is defined as the inner product of wave functions of two ground states
which cannot be accessed directly by the ordinary PIQMC.
Second, a complex-valued weight appears due to the local twist in the interaction.
This causes the ``complex weight problem,'' which is as difficult to deal with as the notorious negative sign problem.
In the present paper, we propose a procedure to calculate the local $Z_2$ Berry phase
by the PIQMC with the loop algorithm~\cite{EvertzLM1993, BeardW1996},
and present some results of the PIQMC simulation for the $S=1/2$ AFH model
on an out-of-phase bond-alternating ladder.
Our method works for any models in arbitrary dimensions as long as the original (untwisted) model
has no negative sign problem.
Extension to other models is straightforward.
We also propose a possibility that the gauge-fixed local Berry connection can be used as another topological order parameter.

First, we define the local Berry phase of AFH models.
To define the local Berry phase at bond a $\left\langle k\ell \right\rangle$,
one twists this bond, that is, replace the term in the Hamiltonian
$S_k^+S_\ell^- + S_k^-S_\ell^+$ by $ e^{i\theta} S_k^+S_\ell^- + e^{-i\theta} S_k^-S_\ell^+$,
where $S^+(S^-)$ are the spin ladder operators and $\theta$ is a twist angle.
The local Berry connection is calculated
as $ A(\theta) = \left\langle \psi(\theta)|\partial_\theta|\psi(\theta)\right\rangle$
by using the ground state $\left.|\psi(\theta)\right\rangle$,
and the Berry phase is then defined as
\begin{equation}
 \gamma = -i \int_0^{2\pi}\!\!\mathrm{d}\theta A(\theta).
 \label{eq:bp}
\end{equation}
To calculate this numerically, the integral needs to be discretized~\cite{KingSmithV1993} as
$ \gamma = \lim_{M\to\infty} \sum_{j=0}^{M-1} \arg \left\langle \psi(\theta_{j}) | \psi(\theta_{j+1})\right\rangle $,
where $\theta_j = 2\pi j/M$.

\begin{figure}[tbp]
\includegraphics[width=.95\linewidth]{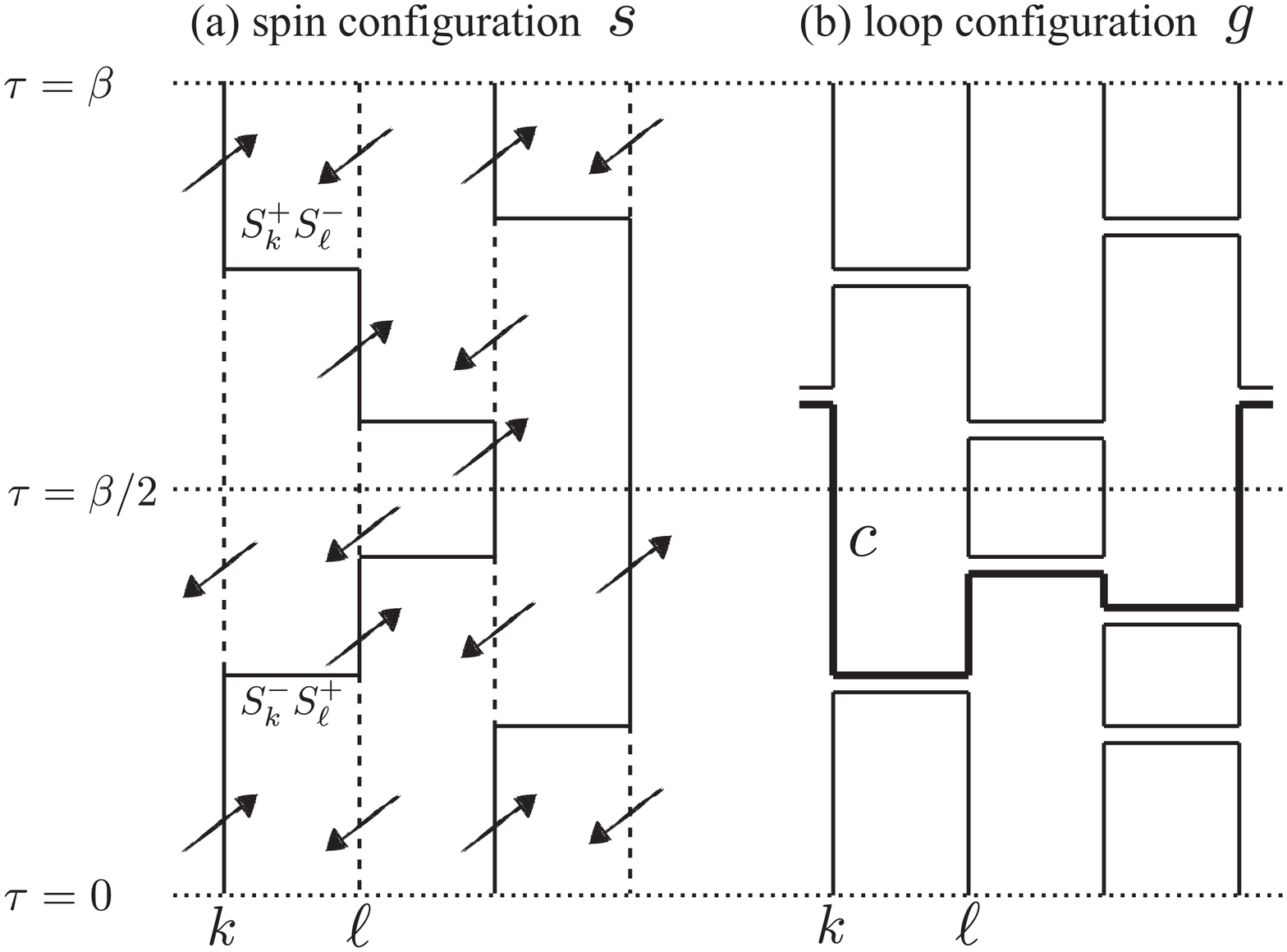}
\caption{
(a) One of the typical worldline configurations of the $S=1/2$ AFH model.
Solid (dashed) vertical lines and upward (downward) arrows denote $S^z=1/2 \,\, (-1/2)$ states.
When one twists the $\langle k\ell \rangle$ bond, $\Ntl = \Nbr = 1$, $\Ntr = \Nbl = 0$.
(b) One of the loop configurations compatible with the spin configuration (a).
The current $n_\pm(c)$ of loop $c$ (thick line) is $n_\pm(c) = (1 - 0 ) \pm (0-0) = 1$ and so this is the meron cluster
when the twist angle $\theta = \pi$.
}
\label{fig:worldline}
\end{figure}

The first obstacle in the PIQMC is how to calculate these inner products.
As explained below, we further deform
the inner product $ B_j \equiv \arg \left\langle \psi(\theta_{j}) | \psi(\theta_{j+1})\right\rangle $
into a Monte Carlo (MC) expectation form: 
$B_j = \arg \sum_\text{s} O(s) W(s) / \sum_\text{s} W(s),$
where $s$ is a configuration index, $O(s)$ is some c-number observable
and $W(s)$ is some positive weight function.
The ground states can be represented by the projection method as
$\left.|\psi(\theta)\right\rangle = \lim_{\beta\to\infty}C \exp\left[-\beta \mathcal{H}(\theta)\right]|\phi(\theta)\rangle$,
where $C$ is a real normalization factor, $|\phi(\theta)\rangle$ is an arbitrary state which is nonorthogonal to the ground state, and $\beta$ is a projection parameter.
Although the $\theta$ dependence of $|\phi(\theta)\rangle$ can be chosen to be arbitrary as long as $|\phi(0)\rangle = |\phi(2\pi)\rangle$,
here we use the classical N\'{e}el state $|\phi\rangle$ for all $\theta$ for simplicity.
The normalization factor $C$ can be neglected since we do not need the absolute value but the argument.
Thus, the projection method leads us to
\begin{equation}
B_j = \arg \left\langle \phi \right|e^{-\frac{\beta}{2}\mathcal{H}(\theta_{j})}e^{-\frac{\beta}{2}\mathcal{H}(\theta_{j+1})}\left|\phi\right\rangle. \label{eq:power}
\end{equation}
As well as the conventional continuous-time PIQMC~\cite{BeardW1996},
we expand the exponential operators by the path integral and
introduce the worldline representation as an MC configuration and a complex valued weight function [see Fig. \ref{fig:worldline} (a)],
by which $B_j$ can be represented as $\arg \sum_s w_j(s).$
Finally, reforming the complex weight to an amplitude and a phase factor,
and dividing by the (real) normalization factor $\sum |w|$
lead us to the following MC form:
\begin{equation}
  B_j = \arg \frac{\sum_s \frac{w_j(s)}{|w_j(s)|}|w_j(s)| }{ \sum_s |w_j(s)|} = \arg \left \langle \frac{w_j}{|w_j|} \right \rangle.
\end{equation}
Note that this deformation is justified since it does not change the argument.

There are two differences from the conventional PIQMC:
First, while the conventional PIQMC has the periodic imaginary-time boundary condition due to the trace operation,
our method has a fixed one due to the gauge fixing~\cite{Hatsugai2004}.
Second, in our simulation the Hamiltonian depends on imaginary time,
$\mathcal{H}(\tau) = \mathcal{H}(\theta_{j+1})$ or $\mathcal{H}(\theta_{j})$,
if $\tau < \beta/2$ or $\tau > \beta/2$, respectively.
Since the effect of twist appears only as the phase of off-diagonal elements of the Hamiltonian,
the phase factor is simply written as 
\begin{equation}
  \frac{w_j}{|w_j|} = \exp\Bigl[i\theta_{j}\left(\Ntr -\Ntl \right) + i\theta_{j+1}\left(\Nbr -\Nbl \right)\Bigr],
\end{equation}
where $\Nbl$ $(\Ntl)$ is the number of off-diagonal operators $S_k^+S_\ell^-$ on the twisted bond $\left\langle k\ell \right\rangle$ at imaginary time $\tau < \beta/2 \quad (\tau > \beta/2)$,
and $\Nbr$ $(\Ntr)$ is the number of off-diagonal operators $S_k^-S_\ell^+$.
Thus, the MC estimator of the Berry phase is obtained as
\begin{equation}
\gamma = \sum_j \arg \Bigl\langle \exp\Bigl[i\theta_{j}\left(\Ntr -\Ntl \right) + i\theta_{j+1}\left(\Nbr -\Nbl \right)\Bigr] \Bigr\rangle_0,
\end{equation}
where $\left\langle\cdots\right\rangle_0$ denotes the MC expectation value with respect to the untwisted system.
Furthermore, we can take the continuous limit $M\to\infty$
by introducing variables $\theta'_j = (\theta_j + \theta_{j+1})/2$ and $ \Delta\theta = \theta_{j+1} - \theta_j= 2\pi/M$:
\begin{equation}
\begin{split}
\gamma &= \sum_j \arg \Bigl\langle \exp\Bigl[i\theta_{j}\left(\Ntr -\Ntl \right) + i\theta_{j+1}\left(\Nbr -\Nbl \right)\Bigr] \Bigr\rangle_0 \\
    &= \sum_j \arg \left\langle e^{i\theta_j' n_+}e^{i\Delta\theta n_- /2} \right\rangle_0 \\
    &\xrightarrow{M\to\infty} \int \arg \left\langle e^{i\theta n_+} \left(1+i\frac{\mathrm{d}\theta}{2}n_-\right)\right\rangle_0 \\
    &= \int \arg \left[ 1 + i \frac{\mathrm{d}\theta}{2} \frac{\left\langle n_-e^{i\theta n_+}\right\rangle_0}{\left\langle e^{i\theta n_+}\right\rangle_0}\right] \\
    &= \int \mathrm{d}\theta \frac{\left\langle n_- e^{i\theta n_+}\right\rangle_0}{2\left\langle e^{i\theta n_+}\right\rangle_0},
\end{split}
\end{equation}
where $n_\pm = (\Nbr - \Nbl) \pm (\Ntr - \Ntl)$
and we used the fact that $\left\langle e^{i\theta n_+}\right\rangle_0$ is real due to imaginary-time reversal symmetry.
By comparing it with the definition of the Berry phase [Eq.~(\ref{eq:bp})], the Berry connection with the present gauge can be calculated by
\begin{equation}
A(\theta) = i \frac{\left\langle n_- e^{in_+\theta}\right\rangle_0}{2\left\langle e^{in_+\theta}\right\rangle_0}.
\label{eq:bc}
\end{equation}

As the projection parameter $\beta$ becomes larger, the width of the distribution of $n_+$ gets broader and so 
both $\left\langle e^{in_+\theta}\right\rangle$ and $\left\langle n_- e^{in_+\theta}\right\rangle$ will be exponentially smaller for $\theta \ne 0$, as $e^{-a\beta \theta^2}$ with some constant $a$.
This means that the relative error of the ratio of these two quantities [Eq.~~(\ref{eq:bc})] will also 
diverge exponentially.
Especially, at $\theta = \pi$ this is nothing but the negative sign problem,
which we commonly encounter for the frustrated AFH model.
The difference from the ordinary fully frustrated models (e.g.\ triangular AFH, kagom\'{e} AFH) is
that the local twist does not introduce the frustration in the diagonal terms but only in the off-diagonal ones,
and this fact fortunately enables us to apply the meron cluster algorithm~\cite{ChandrasekharanW1999},
which is used to solve the negative sign problem for some fermion systems.
Especially, the meron cluster algorithm completely solves the sign problem for the most severe case,
that is, $\theta=\pi$ as seen below.

In the following we will construct the improved estimators~\cite{BrowerCW1998} for the denominator and the numerator of the Berry connection [Eq.~~(\ref{eq:bc})],
and apply the meron algorithm to this problem.
The loop algorithm updates worldline configurations via loop configurations (see Fig.~\ref{fig:worldline}).
We can calculate observables in terms of the loop configuration by tracing out spin variables,
which is referred to as the ``improved estimator.''
First, we define local currents $n_+(c)$ and $n_-(c)$ of each loop $c$ as
\begin{equation}
  n_\pm(c) = [\Nbu(c)-\Nbd(c)] \pm [\Ntu(c)-\Ntd(c)],
\end{equation}
where $\Nbu$ $(\Nbd)$ is the number of downward-upward (upward-downward) turns of the loop on the twisted bond at $\tau < \beta/2$
and $\Ntu$ $(\Ntd)$ is the one at $\tau > \beta/2$.
By using $n_\pm(c)$, we can express the expectation values of the denominator and the numerator in Eq.~~(\ref{eq:bc}) as
\begin{equation}
  \begin{split}
  \left\langle e^{i n_+ \theta} \right\rangle_0
  &= \left\langle \prod_\text{c} \frac{1+e^{i \theta n_+(c)}}{2} \right \rangle_g ,
  \end{split}
\end{equation}
and
\begin{equation}
\begin{split}
  \left\langle n_-e^{i n_+\theta} \right\rangle_0 
= & \left\langle\sum_\text{c}\frac{n_-(c)e^{i\theta n_+(c)}}{2} \prod_{c' \ne c} \frac{1+e^{i\theta n_+(c')}}{2}\right\rangle_g,
\end{split}
\end{equation}
respectively, where $\langle \cdots \rangle_g$ means the average over the loop configurations,
and summation and production are taken over closed loops.
When there are loops with $\theta n_+(c) = \pi \mod 2\pi$,
the contribution to the denominator of Eq.~\ (\ref{eq:bc}) vanishes.
Such loops are called ``meron clusters.''
For example, the loop denoted by the thick line in Fig.~\ref{fig:worldline}(b) has 
$n_+(c) = n_-(c) = 1$, and thus this loop is a meron cluster if $\theta = \pi$.
When there is more than one meron cluster, the numerator of Eq.~~(\ref{eq:bc}) is also zero.
In order to sample loop configurations with a non-zero contribution more frequently
and reduce the statistical error of the Berry connection,
we adopt the histogram reweighting with the Wang-Landau optimization~\cite{WangL2001a}
with respect to the number of the meron clusters.
Since $-iA(\theta)$ is an even function due to the imaginary-time reversal symmetry and has a period $2\pi$,
we fit the numerical data for the discrete values of $\theta$ by $\sum_k a_k \cos(k\theta)$
and obtain the final estimate of the Berry phase as $2\pi a_0$ by integrating the fitting curve.

\begin{figure}[t]
\includegraphics[width=.8\linewidth]{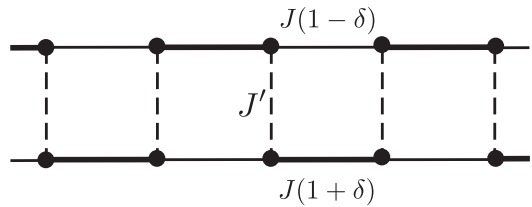}
\caption{
out-of-phase bond-alternating ladder $(0 < \delta < 1)$ .
Singlet pairs are on strong leg bonds (thick lines) or on rung bonds (dashed lines) if $J' < J'_\text{c}(\delta)$ or $J' > J'_\text{c}(\delta)$, respectively. 
}
\label{fig:outphase}
\end{figure}
To demonstrate the present method we calculate the local Berry phase of $S=1/2$ AFH model on an out-of-phase bond-alternating ladder (Fig.~\ref{fig:outphase}).
The program used in the present research was developed based on the ALPS library~\cite{TodoK2001,ALPS2011s}.
The Hamiltonian of the model is
\begin{equation}
\begin{split}
  \mathcal{H} = \sum_{j=1}^{L} [ & J\left(1 + (-1)^{j}\delta\right) \boldsymbol{S}_{1,j}\cdot\boldsymbol{S}_{1,j+1}  \\
                                        +& J\left(1 - (-1)^{j}\delta\right) \boldsymbol{S}_{2,j}\cdot\boldsymbol{S}_{2,j+1} \\
                                        +&  J' \boldsymbol{S}_{1,j}\cdot\boldsymbol{S}_{2,j} ],
\end{split}
\end{equation}
where $L$ is the ladder length (and so the number of sites is $N=2L$) and $\boldsymbol{S}_{i, j}$ stands for the $S=1/2$ spin operator on the $j$-th site of the $i$-th leg.
The boundary condition along the ladder is periodic, that is, $\boldsymbol{S}_{i,L+1} = \boldsymbol{S}_{i,1}$ $(i=1,2)$.
This model exhibits a quantum phase transition at $J' = J'_\text{c}(\delta)$, where the pattern of singlet pairs changes globally;
valence bonds are on strong leg bonds or on rungs for $J' < J'_\text{c}$ or $J' > J'_\text{c}$,
respectively~\cite{MartinDelgadoSS1996}.
\begin{figure}[t]
  \includegraphics[width=\linewidth]{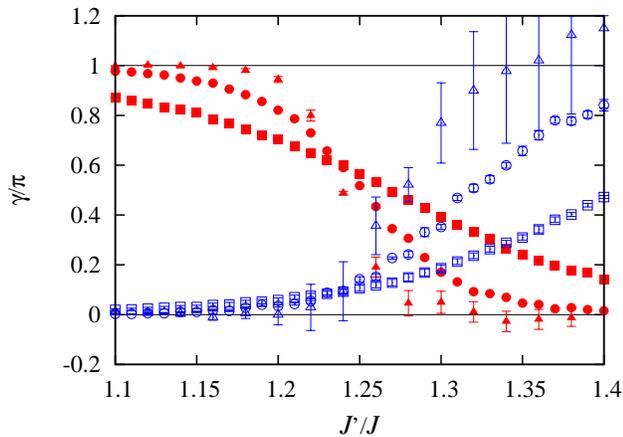}
  \caption{
    (Color online) The local Berry phase on the out-of-phase bond-alternating ladder
    with system size $L=8$ (squares), $16$ (circles), and $32$ (triangles)
    on the leg bonds (solid red symbols) and the rung bonds (open blue symbols).
    The projection parameter $\beta$ is $2L$.
  }
  \label{fig:bp}
\end{figure}
In the present study, we fix $J = 1$ and $\delta = 0.5$,
for which the quantum critical point has been estimated as $J'_\text{c} \sim 1.2$~\cite{Okamoto2003}.
We choose $M = 32$ and the generalized ensemble weight for the loop configuration $g$ as $ w(g,n_m) = w(g)/e^{n_m}$ for $\theta \ne \pi$
while we use the Wang-Landau method for $\theta = \pi$, where $n_m$ is the number of meron clusters.

Figure~\ref{fig:bp} shows the result for system sizes $L=8, 16, 32$ and projection parameter $\beta=2L$.
As $J'/J$ becomes larger, the local Berry phase on the strong leg bonds decreases from $\pi$ to $0$
while the Berry phase on the rung bonds increases from $0$ to $\pi$.
The projection parameter is not large enough to obtain ground states near the critical point
and thus these curves are not step functions.
In this case, however, since the energy gap remains finite except at the critical point, the curves converge to step functions as $L$ and $\beta$ become larger.

\begin{figure}[tb]
\includegraphics[width=\linewidth]{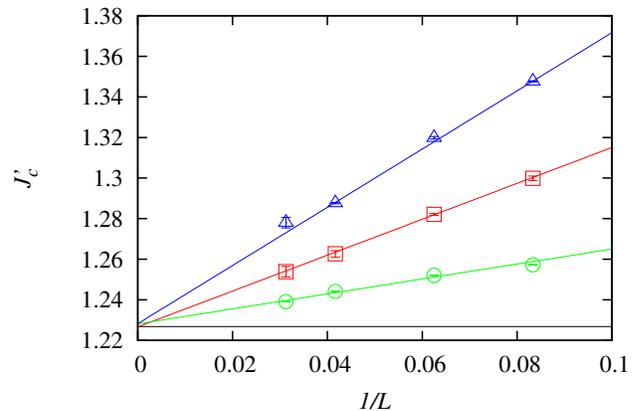}
\caption{
    (Color online) The estimation of the critical point of out-of-phase bond-alternating ladders obtained by the local $Z_2$ Berry phase
    on the leg bond (circles), rung bond (triangles), and their crossing point (squares).
    The horizontal line, $J'_\text{c}/J = 1.2268,$ is the FSS result of staggered susceptibility.
}
\label{fig:criticalpoint}
\end{figure}

We define three points to estimate critical points:
$J_\text{c}^{'\text{leg}}(L)$ and $J_\text{c}^{'\text{rung}}(L)$ are the points where $\gamma^\text{leg} = \pi/2$ and $\gamma^\text{rung} = \pi/2 $, respectively,
and $J_\text{c}^{'\text{cross}}(L)$ is the one where  $\gamma^\text{leg} = \gamma^\text{rung}$.
The critical point in the thermodynamics limit, $J'_\text{c},$ is estimated
by size extrapolation of $J'_\text{c}(L)$ for lattice sizes up to $L=32$;
$J_\text{c}^{'\text{leg}} = 1.2281(18),$ $J_\text{c}^{'\text{rung}} = 1.2282(18)$, and $J_\text{c}^{'\text{cross}} = 1.2266(6)$ (Fig.~\ref{fig:criticalpoint}).
These results are consistent within statistical errors with the independent finite-size scaling (FSS) analysis for the staggered susceptibility, $J'_\text{c}/J = 1.2268(2)$.

\begin{figure}[b]
\includegraphics[width=\linewidth]{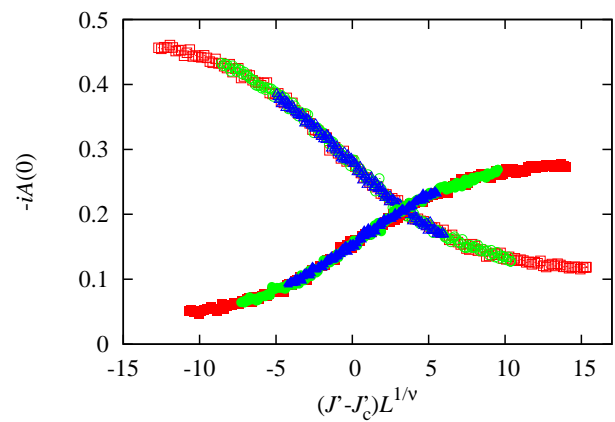}
\caption{
  (Color online) The FSS plot for the imaginary part of the gauge-fixed Berry connection,
  with system sizes $L = 128, 192, 256.$
  The downward to the right curve and the upward one are the leg twist and the rung twist Berry connection, respectively.
}
\label{fig:berryconnection}
\end{figure}
Finally, we point out numerical evidence that the gauge-fixed local Berry connection itself can be a topological order parameter.
In fact, the Berry phase curves for several system sizes cross near the real critical point,
and thus there is no size dependence of the Berry phase by Monte Carlo calculation at the critical point.
We expect that the gauge-fixed local $Z_2$ Berry connection does not also depend on the system size at the critical point
for all twist angles $\theta$.
Note that under the gauge transformation, $|\psi(\theta)\rangle \to e^{i\chi(\theta)} |\psi(\theta)\rangle,$
the Berry connection varies as 
$A(\theta) \to A(\theta) + i\partial_\theta\chi(\theta)$, where $\chi$ is some arbitrary periodic real function such as $\chi(\theta) = \chi(\theta+2\pi)$.
This means that the $J'$-$A$ curve with fixed $\theta$ only shifts vertically by a constant under the gauge transformation
and the crossing point $J'_c$ does not change.
Thus, it is expected that the gauge-fixed Berry connection curves will be on one universal function $f$ by the FSS transformation,
$A(J',N) = f((J'-J'_\text{c})N^{1/\nu})$.
Figure~\ref{fig:berryconnection} shows the result of the Bayesian FSS analysis~\cite{Harada2011}
of the imaginary part of the local Berry connection at $\theta=0$.
The system sizes are $L=128, 192, 256$ and the projection parameter is $\beta = 2L.$
The scaling parameters are 
$J'_\text{c} = 1.2268(1)$ and $\nu = 0.738(8)$ for the leg twist
and $J'_\text{c} = 1.2265(1)$ and $\nu = 0.747(10)$ for the rung twist.
These estimates for the critical point agree with the result of the FSS of staggered susceptibility.
The results of several methods are summarized in TABLE~\ref{table:criticalpoint}.

Calculating the Berry phase for one parameter requires some independent simulations.
In order to suppress the complex weight problem one has to use the reweighting method for the number of meron clusters,
which is a global variable and so the simulation is heavy and difficult to be parallelized.
The local Berry connection for $\theta=0$, on the other hand, can be calculated from one simulation
without the meron cluster algorithm and the reweighting method,
and thus it is easy to calculate it and at the same time simulation can be parallelized.

In conclusion, we presented the procedure to calculate the local $Z_2$ Berry phase by the PIQMC
and also proposed that the gauge-fixed local Berry connection can be used as another topological order parameter.
We calculated these order parameters of the AFH ladder models for a demonstration,
and the result for the thermodynamics limit are consistent with the result attained by conventional methods.

\begin{table}[tb]
  \caption{
    The quantum critical point $J'_c/J$ of out-of-phase bond-alternating ladder with dimerization $\delta=0.5$ obtained by each method.
  }
  \label{table:criticalpoint}
  \begingroup
  \renewcommand{\arraystretch}{1.5}
  \begin{tabular}{lcc}
    &&\\
    \hline\hline
    method & $L_{\text{max}}$ & $J'_\text{c} / J$ \\
    \hline
    Berry phase (leg twist)       & $32$  & $1.2281(18)$ \\
    Berry phase (rung twist)      & $32$  & $1.2282(18)$ \\
    Berry phase (cross)           & $32$  & $1.2266(6)$  \\
    Berry connection (leg twist)  & $256$ & $1.2268(1)$  \\
    Berry connection (rung twist) & $256$ & $1.2265(1)$  \\
    FSS of susceptibility         & $256$ & $1.2268(2)$  \\
    \hline\hline
    &&
  \end{tabular}
  \endgroup
\end{table}

We acknowledge support by
KAKENHI (No.~23540438),
JSPS,
Grand Challenges in Next-Generation Integrated Nanoscience,
Next-Generation Supercomputer Project,
the HPCI Strategic Programs for Innovative Research (SPIRE),
the Global COE program ``the Physical Sciences Frontier,''
MEXT, Japan,
and the Computational Materials Science Initiative (CMSI).
\bibliography{bp}

\begin{thebibliography}{26}%
\makeatletter
\providecommand \@ifxundefined [1]{%
 \@ifx{#1\undefined}
}%
\providecommand \@ifnum [1]{%
 \ifnum #1\expandafter \@firstoftwo
 \else \expandafter \@secondoftwo
 \fi
}%
\providecommand \@ifx [1]{%
 \ifx #1\expandafter \@firstoftwo
 \else \expandafter \@secondoftwo
 \fi
}%
\providecommand \natexlab [1]{#1}%
\providecommand \enquote  [1]{``#1''}%
\providecommand \bibnamefont  [1]{#1}%
\providecommand \bibfnamefont [1]{#1}%
\providecommand \citenamefont [1]{#1}%
\providecommand \href@noop [0]{\@secondoftwo}%
\providecommand \href [0]{\begingroup \@sanitize@url \@href}%
\providecommand \@href[1]{\@@startlink{#1}\@@href}%
\providecommand \@@href[1]{\endgroup#1\@@endlink}%
\providecommand \@sanitize@url [0]{\catcode `\\12\catcode `\$12\catcode
  `\&12\catcode `\#12\catcode `\^12\catcode `\_12\catcode `\%12\relax}%
\providecommand \@@startlink[1]{}%
\providecommand \@@endlink[0]{}%
\providecommand \url  [0]{\begingroup\@sanitize@url \@url }%
\providecommand \@url [1]{\endgroup\@href {#1}{\urlprefix }}%
\providecommand \urlprefix  [0]{URL }%
\providecommand \Eprint [0]{\href }%
\providecommand \doibase [0]{http://dx.doi.org/}%
\providecommand \selectlanguage [0]{\@gobble}%
\providecommand \bibinfo  [0]{\@secondoftwo}%
\providecommand \bibfield  [0]{\@secondoftwo}%
\providecommand \translation [1]{[#1]}%
\providecommand \BibitemOpen [0]{}%
\providecommand \bibitemStop [0]{}%
\providecommand \bibitemNoStop [0]{.\EOS\space}%
\providecommand \EOS [0]{\spacefactor3000\relax}%
\providecommand \BibitemShut  [1]{\csname bibitem#1\endcsname}%
\let\auto@bib@innerbib\@empty
\bibitem [{\citenamefont {Haldane}(1982)}]{Haldane1982}%
  \BibitemOpen
  \bibfield  {author} {\bibinfo {author} {\bibfnamefont {F.~D.~M.}\
  \bibnamefont {Haldane}},\ }\href@noop {} {\bibfield  {journal} {\bibinfo
  {journal} {Phys. Rev. B}\ }\textbf {\bibinfo {volume} {25}},\ \bibinfo
  {pages} {4925} (\bibinfo {year} {1982})}\BibitemShut {NoStop}%
\bibitem [{\citenamefont {Affleck}\ \emph {et~al.}(1987)\citenamefont
  {Affleck}, \citenamefont {Kennedy}, \citenamefont {Lieb},\ and\ \citenamefont
  {Tasaki}}]{AffleckKLT1987}%
  \BibitemOpen
  \bibfield  {author} {\bibinfo {author} {\bibfnamefont {I.}~\bibnamefont
  {Affleck}}, \bibinfo {author} {\bibfnamefont {T.}~\bibnamefont {Kennedy}},
  \bibinfo {author} {\bibfnamefont {E.~H.}\ \bibnamefont {Lieb}}, \ and\
  \bibinfo {author} {\bibfnamefont {H.}~\bibnamefont {Tasaki}},\ }\href@noop {}
  {\bibfield  {journal} {\bibinfo  {journal} {Phys. Rev. Lett.}\ }\textbf
  {\bibinfo {volume} {59}},\ \bibinfo {pages} {799} (\bibinfo {year}
  {1987})}\BibitemShut {NoStop}%
\bibitem [{\citenamefont {den Nijs}\ and\ \citenamefont
  {Rommelse}(1989)}]{NijsR1989}%
  \BibitemOpen
  \bibfield  {author} {\bibinfo {author} {\bibfnamefont {M.}~\bibnamefont {den
  Nijs}}\ and\ \bibinfo {author} {\bibfnamefont {K.}~\bibnamefont {Rommelse}},\
  }\href@noop {} {\bibfield  {journal} {\bibinfo  {journal} {Phys. Rev. B}\
  }\textbf {\bibinfo {volume} {40}},\ \bibinfo {pages} {4709} (\bibinfo {year}
  {1989})}\BibitemShut {NoStop}%
\bibitem [{\citenamefont {Oshikawa}(1992)}]{Oshikawa1992}%
  \BibitemOpen
  \bibfield  {author} {\bibinfo {author} {\bibfnamefont {M.}~\bibnamefont
  {Oshikawa}},\ }\href@noop {} {\bibfield  {journal} {\bibinfo  {journal} {J.
  Phys. Condens. Matter}\ }\textbf {\bibinfo {volume} {4}},\ \bibinfo {pages}
  {7469} (\bibinfo {year} {1992})}\BibitemShut {NoStop}%
\bibitem [{\citenamefont {Todo}\ \emph {et~al.}(2001)\citenamefont {Todo},
  \citenamefont {Matsumoto}, \citenamefont {Yasuda},\ and\ \citenamefont
  {Takayama}}]{TodoMYT2001}%
  \BibitemOpen
  \bibfield  {author} {\bibinfo {author} {\bibfnamefont {S.}~\bibnamefont
  {Todo}}, \bibinfo {author} {\bibfnamefont {M.}~\bibnamefont {Matsumoto}},
  \bibinfo {author} {\bibfnamefont {C.}~\bibnamefont {Yasuda}}, \ and\ \bibinfo
  {author} {\bibfnamefont {H.}~\bibnamefont {Takayama}},\ }\href@noop {}
  {\bibfield  {journal} {\bibinfo  {journal} {Phys. Rev. B}\ }\textbf {\bibinfo
  {volume} {64}},\ \bibinfo {pages} {224412} (\bibinfo {year}
  {2001})}\BibitemShut {NoStop}%
\bibitem [{\citenamefont {Nakamura}\ and\ \citenamefont
  {Todo}(2002)}]{NakamuraT2002a}%
  \BibitemOpen
  \bibfield  {author} {\bibinfo {author} {\bibfnamefont {M.}~\bibnamefont
  {Nakamura}}\ and\ \bibinfo {author} {\bibfnamefont {S.}~\bibnamefont
  {Todo}},\ }\href@noop {} {\bibfield  {journal} {\bibinfo  {journal} {Phys.
  Rev. Lett.}\ }\textbf {\bibinfo {volume} {89}},\ \bibinfo {pages} {077204}
  (\bibinfo {year} {2002})}\BibitemShut {NoStop}%
\bibitem [{\citenamefont {Hatsugai}(2006)}]{Hatsugai2006}%
  \BibitemOpen
  \bibfield  {author} {\bibinfo {author} {\bibfnamefont {Y.}~\bibnamefont
  {Hatsugai}},\ }\href@noop {} {\bibfield  {journal} {\bibinfo  {journal} {J.
  Phys. Soc. Jpn.}\ }\textbf {\bibinfo {volume} {75}},\ \bibinfo {pages}
  {123601} (\bibinfo {year} {2006})}\BibitemShut {NoStop}%
\bibitem [{\citenamefont {Maruyama}\ and\ \citenamefont
  {Hatsugai}(2007)}]{MaruyamaH2007}%
  \BibitemOpen
  \bibfield  {author} {\bibinfo {author} {\bibfnamefont {I.}~\bibnamefont
  {Maruyama}}\ and\ \bibinfo {author} {\bibfnamefont {Y.}~\bibnamefont
  {Hatsugai}},\ }\href@noop {} {\bibfield  {journal} {\bibinfo  {journal} {J.
  Phys. Soc. Jpn.}\ }\textbf {\bibinfo {volume} {76}},\ \bibinfo {pages}
  {113601} (\bibinfo {year} {2007})}\BibitemShut {NoStop}%
\bibitem [{\citenamefont {Maruyama}\ and\ \citenamefont
  {Hatsugai}(2009)}]{MaruyamaH2009}%
  \BibitemOpen
  \bibfield  {author} {\bibinfo {author} {\bibfnamefont {I.}~\bibnamefont
  {Maruyama}}\ and\ \bibinfo {author} {\bibfnamefont {Y.}~\bibnamefont
  {Hatsugai}},\ }\href@noop {} {\bibfield  {journal} {\bibinfo  {journal}
  {Journal of Physics: Conference Series}\ }\textbf {\bibinfo {volume} {150}},\
  \bibinfo {pages} {042116} (\bibinfo {year} {2009})}\BibitemShut {NoStop}%
\bibitem [{\citenamefont {Ryu}\ and\ \citenamefont
  {Hatsugai}(2006)}]{RyuH2006}%
  \BibitemOpen
  \bibfield  {author} {\bibinfo {author} {\bibfnamefont {S.}~\bibnamefont
  {Ryu}}\ and\ \bibinfo {author} {\bibfnamefont {Y.}~\bibnamefont {Hatsugai}},\
  }\href@noop {} {\bibfield  {journal} {\bibinfo  {journal} {Phys. Rev. B}\
  }\textbf {\bibinfo {volume} {73}},\ \bibinfo {pages} {245115} (\bibinfo
  {year} {2006})}\BibitemShut {NoStop}%
\bibitem [{\citenamefont {Hirano}\ \emph {et~al.}(2008)\citenamefont {Hirano},
  \citenamefont {Katsura},\ and\ \citenamefont {Hatsugai}}]{HiranoKH2008}%
  \BibitemOpen
  \bibfield  {author} {\bibinfo {author} {\bibfnamefont {T.}~\bibnamefont
  {Hirano}}, \bibinfo {author} {\bibfnamefont {H.}~\bibnamefont {Katsura}}, \
  and\ \bibinfo {author} {\bibfnamefont {Y.}~\bibnamefont {Hatsugai}},\
  }\href@noop {} {\bibfield  {journal} {\bibinfo  {journal} {Phys. Rev. B}\
  }\textbf {\bibinfo {volume} {77}},\ \bibinfo {pages} {094431} (\bibinfo
  {year} {2008})}\BibitemShut {NoStop}%
\bibitem [{\citenamefont {Kitaev}\ and\ \citenamefont
  {Preskill}(2006)}]{KitaevP2006}%
  \BibitemOpen
  \bibfield  {author} {\bibinfo {author} {\bibfnamefont {A.}~\bibnamefont
  {Kitaev}}\ and\ \bibinfo {author} {\bibfnamefont {J.}~\bibnamefont
  {Preskill}},\ }\href@noop {} {\bibfield  {journal} {\bibinfo  {journal}
  {Phys. Rev. Lett.}\ }\textbf {\bibinfo {volume} {96}},\ \bibinfo {pages}
  {110404} (\bibinfo {year} {2006})}\BibitemShut {NoStop}%
\bibitem [{\citenamefont {Levin}\ and\ \citenamefont {Wen}(2006)}]{LevinW2006}%
  \BibitemOpen
  \bibfield  {author} {\bibinfo {author} {\bibfnamefont {M.}~\bibnamefont
  {Levin}}\ and\ \bibinfo {author} {\bibfnamefont {X.-G.}\ \bibnamefont
  {Wen}},\ }\href@noop {} {\bibfield  {journal} {\bibinfo  {journal} {Phys.
  Rev. Lett.}\ }\textbf {\bibinfo {volume} {96}},\ \bibinfo {pages} {110405}
  (\bibinfo {year} {2006})}\BibitemShut {NoStop}%
\bibitem [{\citenamefont {Suzuki}(1976)}]{Suzuki1976}%
  \BibitemOpen
  \bibfield  {author} {\bibinfo {author} {\bibfnamefont {M.}~\bibnamefont
  {Suzuki}},\ }\href@noop {} {\bibfield  {journal} {\bibinfo  {journal} {Prog.
  Theor. Phys.}\ }\textbf {\bibinfo {volume} {56}},\ \bibinfo {pages} {1454}
  (\bibinfo {year} {1976})}\BibitemShut {NoStop}%
\bibitem [{\citenamefont {Evertz}\ \emph {et~al.}(1993)\citenamefont {Evertz},
  \citenamefont {Lana},\ and\ \citenamefont {Marcu}}]{EvertzLM1993}%
  \BibitemOpen
  \bibfield  {author} {\bibinfo {author} {\bibfnamefont {H.~G.}\ \bibnamefont
  {Evertz}}, \bibinfo {author} {\bibfnamefont {G.}~\bibnamefont {Lana}}, \ and\
  \bibinfo {author} {\bibfnamefont {M.}~\bibnamefont {Marcu}},\ }\href@noop {}
  {\bibfield  {journal} {\bibinfo  {journal} {Phys. Rev. Lett.}\ }\textbf
  {\bibinfo {volume} {70}},\ \bibinfo {pages} {875} (\bibinfo {year}
  {1993})}\BibitemShut {NoStop}%
\bibitem [{\citenamefont {Beard}\ and\ \citenamefont
  {Wiese}(1996)}]{BeardW1996}%
  \BibitemOpen
  \bibfield  {author} {\bibinfo {author} {\bibfnamefont {B.~B.}\ \bibnamefont
  {Beard}}\ and\ \bibinfo {author} {\bibfnamefont {U.~J.}\ \bibnamefont
  {Wiese}},\ }\href@noop {} {\bibfield  {journal} {\bibinfo  {journal} {Phys.
  Rev. Lett.}\ }\textbf {\bibinfo {volume} {77}},\ \bibinfo {pages} {5130}
  (\bibinfo {year} {1996})}\BibitemShut {NoStop}%
\bibitem [{\citenamefont {King-Smith}\ and\ \citenamefont
  {Vanderbilt}(1993)}]{KingSmithV1993}%
  \BibitemOpen
  \bibfield  {author} {\bibinfo {author} {\bibfnamefont {R.~D.}\ \bibnamefont
  {King-Smith}}\ and\ \bibinfo {author} {\bibfnamefont {D.}~\bibnamefont
  {Vanderbilt}},\ }\href@noop {} {\bibfield  {journal} {\bibinfo  {journal}
  {Phys. Rev. B}\ }\textbf {\bibinfo {volume} {47}},\ \bibinfo {pages} {1651}
  (\bibinfo {year} {1993})}\BibitemShut {NoStop}%
\bibitem [{\citenamefont {Hatsugai}(2004)}]{Hatsugai2004}%
  \BibitemOpen
  \bibfield  {author} {\bibinfo {author} {\bibfnamefont {Y.}~\bibnamefont
  {Hatsugai}},\ }\href@noop {} {\bibfield  {journal} {\bibinfo  {journal} {J.
  Phys. Soc. Jpn.}\ }\textbf {\bibinfo {volume} {73}},\ \bibinfo {pages} {2604}
  (\bibinfo {year} {2004})}\BibitemShut {NoStop}%
\bibitem [{\citenamefont {Chandrasekharan}\ and\ \citenamefont
  {Wiese}(1999)}]{ChandrasekharanW1999}%
  \BibitemOpen
  \bibfield  {author} {\bibinfo {author} {\bibfnamefont {S.}~\bibnamefont
  {Chandrasekharan}}\ and\ \bibinfo {author} {\bibfnamefont {U.-J.}\
  \bibnamefont {Wiese}},\ }\href@noop {} {\bibfield  {journal} {\bibinfo
  {journal} {Phys. Rev. Lett.}\ }\textbf {\bibinfo {volume} {83}},\ \bibinfo
  {pages} {3116} (\bibinfo {year} {1999})}\BibitemShut {NoStop}%
\bibitem [{\citenamefont {Brower}\ \emph {et~al.}(1998)\citenamefont {Brower},
  \citenamefont {Chandrasekharan},\ and\ \citenamefont {Wiese}}]{BrowerCW1998}%
  \BibitemOpen
  \bibfield  {author} {\bibinfo {author} {\bibfnamefont {R.}~\bibnamefont
  {Brower}}, \bibinfo {author} {\bibfnamefont {S.}~\bibnamefont
  {Chandrasekharan}}, \ and\ \bibinfo {author} {\bibfnamefont {U.-J.}\
  \bibnamefont {Wiese}},\ }\href@noop {} {\bibfield  {journal} {\bibinfo
  {journal} {Physica A}\ }\textbf {\bibinfo {volume} {261}},\ \bibinfo {pages}
  {520} (\bibinfo {year} {1998})}\BibitemShut {NoStop}%
\bibitem [{\citenamefont {Wang}\ and\ \citenamefont
  {Landau}(2001)}]{WangL2001a}%
  \BibitemOpen
  \bibfield  {author} {\bibinfo {author} {\bibfnamefont {F.}~\bibnamefont
  {Wang}}\ and\ \bibinfo {author} {\bibfnamefont {D.~P.}\ \bibnamefont
  {Landau}},\ }\href@noop {} {\bibfield  {journal} {\bibinfo  {journal} {Phys.
  Rev. Lett.}\ }\textbf {\bibinfo {volume} {86}},\ \bibinfo {pages} {2050}
  (\bibinfo {year} {2001})}\BibitemShut {NoStop}%
\bibitem [{\citenamefont {Todo}\ and\ \citenamefont {Kato}(2001)}]{TodoK2001}%
  \BibitemOpen
  \bibfield  {author} {\bibinfo {author} {\bibfnamefont {S.}~\bibnamefont
  {Todo}}\ and\ \bibinfo {author} {\bibfnamefont {K.}~\bibnamefont {Kato}},\
  }\href@noop {} {\bibfield  {journal} {\bibinfo  {journal} {Phys. Rev. Lett.}\
  }\textbf {\bibinfo {volume} {87}},\ \bibinfo {pages} {047203} (\bibinfo
  {year} {2001})}\BibitemShut {NoStop}%
\bibitem [{\citenamefont {Bauer}\ and\ \citenamefont {{\it et
  al.}}(2011)}]{ALPS2011s}%
  \BibitemOpen
  \bibfield  {author} {\bibinfo {author} {\bibfnamefont {B.}~\bibnamefont
  {Bauer}}\ and\ \bibinfo {author} {\bibnamefont {{\it et al.}}},\ }\href@noop
  {} {\bibfield  {journal} {\bibinfo  {journal} {J. Stat Mech.}\ ,\ \bibinfo
  {pages} {P05001}} (\bibinfo {year} {2011})}\BibitemShut {NoStop}%
\bibitem [{\citenamefont {Mart\'in-Delgado}\ \emph {et~al.}(1996)\citenamefont
  {Mart\'in-Delgado}, \citenamefont {Shankar},\ and\ \citenamefont
  {Sierra}}]{MartinDelgadoSS1996}%
  \BibitemOpen
  \bibfield  {author} {\bibinfo {author} {\bibfnamefont {M.~A.}\ \bibnamefont
  {Mart\'in-Delgado}}, \bibinfo {author} {\bibfnamefont {R.}~\bibnamefont
  {Shankar}}, \ and\ \bibinfo {author} {\bibfnamefont {G.}~\bibnamefont
  {Sierra}},\ }\href@noop {} {\bibfield  {journal} {\bibinfo  {journal} {Phys.
  Rev. Lett.}\ }\textbf {\bibinfo {volume} {77}},\ \bibinfo {pages} {3443}
  (\bibinfo {year} {1996})}\BibitemShut {NoStop}%
\bibitem [{\citenamefont {Okamoto}(2003)}]{Okamoto2003}%
  \BibitemOpen
  \bibfield  {author} {\bibinfo {author} {\bibfnamefont {K.}~\bibnamefont
  {Okamoto}},\ }\href@noop {} {\bibfield  {journal} {\bibinfo  {journal} {Phys.
  Rev. B}\ }\textbf {\bibinfo {volume} {67}},\ \bibinfo {pages} {212408}
  (\bibinfo {year} {2003})}\BibitemShut {NoStop}%
\bibitem [{\citenamefont {Harada}(2011)}]{Harada2011}%
  \BibitemOpen
  \bibfield  {author} {\bibinfo {author} {\bibfnamefont {K.}~\bibnamefont
  {Harada}},\ }\href@noop {} {\bibfield  {journal} {\bibinfo  {journal} {Phys.
  Rev. E}\ }\textbf {\bibinfo {volume} {84}},\ \bibinfo {pages} {056704}
  (\bibinfo {year} {2011})}\BibitemShut {NoStop}%
\end{thebibliography}%
\end{document}